\documentclass[11pt,a4paper]{article}

\usepackage{graphicx}                           
\usepackage{amsmath, amssymb, amsthm}           
\usepackage[T1]{url}                            
\usepackage{url}                                
\usepackage{tikz}
\usetikzlibrary{quantikz2}
\usepackage{pgfgantt}
\usepackage{rotating}

\usepackage[left=2cm,top=1.5cm,bottom=1.5cm,right=2cm]{geometry}    
\usepackage{setspace}                                               
\usepackage{parskip}                                                
\numberwithin{equation}{section}                
\usepackage[%
style=phys,%
biblabel=brackets,%
doi=false,
minnames=3,
maxnames=99,
articletitle=true
]
{biblatex}                  
\AtEveryBibitem{\clearlist{language}}
\def\biblio{\clearpage\printbibliography} 

\usepackage{hyperref}   

\title{\textbf{A map of indefinite causal order}}
\author{Jorge Escand\'on-Monardes\thanks{jescandon@udec.cl}\\Millennium Institute for Research in Optics, Universidad de Concepci\'on, Chile\\Departamento de F\'isica, Universidad de Concepci\'on, 160-C Concepci\'on, Chile}
\date{June 2025}

\begin{document}
\def\biblio{}   

\maketitle

\section*{Prologue}

This document started as an excerpt of the PhD Thesis by Jorge Escandón, entitled ``Indefinite causal order towards continuous-variable quantum systems''. It is intended as a map of indefinite causal order, which can be used to guide the trip of any explorer in the field. It offers an overview of the different approaches and the main ongoing debates in an organised way. But a map never replace the terrain, so it cannot be considered as an introduction to the field. All the tools required for a good trip are developed in the corresponding literature; they depend on the approach and the specific problems to be addressed. The map is written at a conceptual level, without mathematical technicalities while full of references. Anyone who is new to the field can use the map as a first sight before deciding where to start their journey, according to their own interest and background. Any experienced researcher can use the map to get quick access to all relevant literature in any of the threads related to indefinite causal order. The list of references is as complete as possible and is being updated continuously. We apologise for any work that may have been omitted and gently ask the reader to provide new references that they could consider relevant to the field.

\section{Motivation}

Let me invite you to listen to your favourite song. Please, look for it and press play. Looking at this map can wait a few minutes.

Ready? Cool!

If you really played your favourite song, I could guess that you did not skip any part of it, you started listening from the first second of the recording until the last one. No pauses, no steps back and forth. A well defined sequence of sounds in a specific rhythm, carefully studied and performed by the artist. Now look for the album to which that song belongs. The order of the tracks in the album is most likely carefully designed as well, regardless you could choose the order in which you want to listen to them. If you want, you could randomly reshuffle the order of the tracks, although the whole experience would be a little different. Yes, you can reshuffle your playlist. But can you reshuffle the episodes of your life? Can you reshuffle the order of physical phenomena?

In Nature, some things cannot happen if another event, its cause, has not occurred beforehand. For example, we cannot expect a water leakage from a pipeline before the pipeline breaks, nor a plane to fly before being built. The effects cannot precede their causes. The order of the events is well defined and arranged in a timeline that runs from past to future, as in the best of musical compositions. We say that physical events are sorted in a \textit{definite causal order}. By the way, the arrow of time is like a stubborn leader who does not wait nor turn back for us. We better give up trying to pause or remix the song of Nature!

Nevertheless, here we deal with \textit{indefinite causal order} (ICO). There are good reasons to explore the possibility of events without a definite order. The first of them is to examine the boundaries of the theories that we use to describe the universe. The second reason is plain: quantum mechanics allows for it.

In this ``map'' we are going to present the concept of indefinite causal order and make a quick journey through its different flavours. We will start with a broad conceptual motivation for studying ICO, based on the approach of Lucien Hardy to quantum gravity. Then, we will introduce the quantum switch, which is a particular instance of ICO, followed by the process matrices framework and the gravitational quantum switch. Finally, we will comment on some ongoing debates within the field. 

Now let us start the trip and let the arrow of time flow.

\section{Hardy's program}

Quantum theory and general relativity are the most succesful physical theories so far. They are able to explain together most of known physical phenomena. However, the golden dream of several physicists is to explain everything in one single theory involving both of them, that is, to achieve a sort of ``quantum gravity'' theory. The search for such a theory has been the motive for many researchers, Lucien Hardy among them. In 2007, introducing a very interesting approach, Hardy highligthed that general relativity is a deterministic theory where the causal structure is not predefined (it depends on the distribution of matter and energy), while quantum theory is a fundamentally probabilistic theory on a fixed causal structure\footnote{The causal structure of spacetime is usually understood as an order relation between spacetime points. Two points can be causally connected only if it is possible to send a signal from one to the other without surpassing the speed of light. We could say that the causal structure is a feature of the manifold describing spacetime, hence it depends on the distribution of matter and energy. In the case of quantum theory, we usually assume a fixed flat geometry, since quantum experiments are performed in a lab within a locally Minkowski spacetime.}. According to his reasoning, a quantum gravity theory should be expected to combine the radical aspects of both theories. In other words, it should be a \textit{probabilistic theory with non-fixed causal structure} \cite{Hardy_2007}.

In a subsequent paper \cite{Hardy_2009}, Hardy discussed the consequences of an indefinite causal structure for computation. Algorithms, which are specific sequences of calculations, are the basic elements for computing. How could any algorithm be implemented or even defined when the causal structure and the order of events is non-fixed? Can an eventual quantum gravity computer outperform quantum and classical computers? By raising these questions Hardy implicitly invited the community of quantum information and quantum computation to join his research program.

Inspired by Hardy's ideas, at least three different approaches to indefinite causal order have been developed:
\begin{itemize}
    \item \textbf{Quantum switch:} The first approach is based solely on quantum computation. It defines a higher-order operation, called the \textit{quantum switch}, which consists in the application of two or more quantum operations in a coherent superposition of different orders, controlled by an ancillary system. The order of the operations remains indefinite unless the control system is measured in the computational basis. The quantum switch is a process allowed by quantum mechanics and has been already implemented in a number of experiments.
    \item \textbf{Process matrices framework:} In this approach, two parties are free to implement quantum operations on a quantum system according to the laws of quantum mechanics. Inside their labs there is a fixed Minkowski spacetime, but no causal structure is assumed out of the labs. Here, the connection between the outcomes and inputs of each lab is mediated by a \textit{process matrix}. Physical processes such as state preparation, measurements and signalling from one lab to another can be fairly described by particular process matrices, but some non-physical processes are also allowed. Therefore, the process matrices framework goes beyond quantum mechanics, although without dealing with gravity explicitly. One important feature of this approach is a precise definition of the notion of \textit{causal nonseparability}, which is required to decide whether a given process is an instance of indefintie causal order or not. Actually, it has been shown, both theoretically and experimentally, that the quantum switch is a causally nonseparable process in this sense.
    \item \textbf{Superposition of causal structures:} The last approach is one that explicitly takes gravity into account. It postulates that spacetime can be treated as a quantum system and therefore can be assigned a quantum state. Particularly, any classical manifold can be related to a specific quantum state of the spacetime. Hence, an indefinite causal structure can be achieved just by superposing different manifolds. Interestingly, some particular gravitational scenarios can be devised in such a way that they reproduce the same effects as the quantum switch. This scenarios are naturally known as \textit{gravitational quantum switches}.
\end{itemize}

In the following sections we review these three different approaches to indefinite causal order. Although they raise relevant concerns regarding their physicality (see Section \ref{sec1_1_sub_Debates} for an overview on some ongoing debates), they push forward our understanding of quantum mechanics and explore what kind of phenomena we could expect in the realm of quantum gravity.

\section{The quantum switch}

\subsection{Concept}

\begin{figure}[t]
    \centering
    \includegraphics[width=0.8\linewidth]{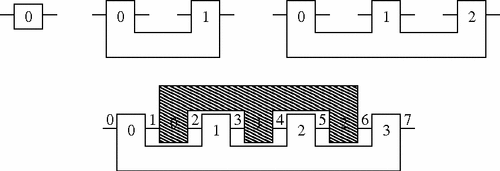}    
    \caption[Quantum supermaps and quantum combs.]{\textbf{Quantum supermaps and quantum combs.} A box represents a quantum operation or $1$-comb (upper left); a diagram with two teeth represents a $2$-comb, i.e., a supermap where a quantum box can be plugged (upper center); a diagram with three teeths represents a $3$-comb where up to two boxes or one $2$-comb can be plugged (upper right). When two combs are combined, the result is a new comb. For example, a $4$-comb can take a $3$-comb as input and transform it in a 1-comb as output (lower). Figure from Ref. \cite{Chiribella_2009}.}
    \label{Fig_1_Qcombs}
\end{figure}

The usual way to proceed in quantum computing is by encoding information in the state of a quantum system and then to process that information by performing certain quantum operations on the system. The final state of the quantum system is measured and hence the intended output of the computation is retrieved. This procedure is analogous to the evaluation of a function on a given input. But just like higher-order functions take other functions as input or output, we could define transformations which map quantum operations into quantum operations. Such higher-order maps are known as quantum supermaps and were introduced in Ref. \cite{Chiribella_2008} and extended to the notion of quantum combs in Ref. \cite{Chiribella_2009}. Operationally, these transformations can be seen as circuits with slots where the input operations, provided as black boxes or oracles, can be plugged. The resulting circuit corresponds to the output quantum operation (see Fig. \ref{Fig_1_Qcombs}).

\begin{figure}[t]
    \centering
    \includegraphics[width=0.6\linewidth]{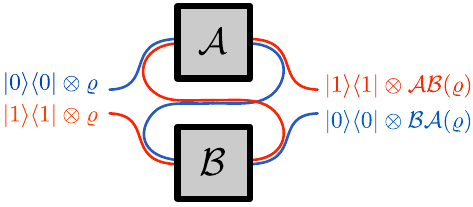}    
    \caption[Quantum switch.]{\textbf{Quantum switch.} The simplest quantum switch applies two gates in two different orders, which is coherently controlled by a control qubit. In photonic realizations, the order of the operations can be controlled by the path degree of freedom of photons, which we illustrate by colour wires. When the control is in a superposition of states, the order of the gates becomes indefinite.}
    \label{Fig_1_Qswitch}
\end{figure}

The quantum switch, introduced in 2013 by Chiribella et al. \cite{Chiribella_2013}, is a special example of quantum supermap. It consists in the application of two quantum channels $\mathcal{A}$ and $\mathcal{B}$ on a quantum system in one of two different orders, either $\mathcal{B\circ A}$ or $\mathcal{A\circ B}$. The order is coherently controlled by an ancillary qubit. Hereafter, the system on which $\mathcal{A}$ and $\mathcal{B}$ act will be called the \textit{target system}, while the ancillary qubit will be called the \textit{control system}. If the control is in state $\ket{0}$, $\mathcal{A}$ is applied before $\mathcal{B}$; conversely, if the control is in state $\ket{1}$, $\mathcal{B}$ is applied before $\mathcal{A}$ (see Fig. \ref{Fig_1_Qswitch}). Notice that this is a straightforward extension of a classical code where two subroutines are called in one specific order which depends on the truth value of an \texttt{if} statement. Here, the novelty is that the control system is characterized by a quantum state, which could be a superposition of $\ket{0}$ and $\ket{1}$, leading to a ``superposition'' of two different orders. For example, if the state of the control system is $\ket{+}=(\ket{0}+\ket{1})/\sqrt{2}$, then the order of the operations will remain indefinite even after the application of both operations. Indeed, the order can be fixed in a later time by measuring the control system in the computational basis, with 50\% of probability of getting one order or the other. Furthermore, the control system could be measured in a different basis, such as the $\sigma_x$ eigenbasis, leading to an effective transformation on the target system that is not one of the two orders anymore. Particularly, if $\mathcal{A}$ and $\mathcal{B}$ are unitary operations, the target system will have undergone either the commutator or anticommutator of these operations.

\begin{figure}[t]
    \centering
    \begin{quantikz}[row sep={0.7cm,between origins}, column sep={0.4cm}]
        \lstick{$|+\rangle$} & \ctrl{2} & \qw & \ctrl{2} & \qw & \octrl{2} & \qw & \octrl{2} & \qw\rstick[wires=2]{$\frac{1}{\sqrt{2}}\big(\ket{0}\otimes BA\ket{\psi}+\ket{1}\otimes AB\ket{\psi}\big)$}
        \\
        \lstick{$|\psi\rangle$} & \targX{} & \gate{A} & \targX{} & \gate{B} & \targX{} & \gate{A} & \targX{} & \qw
        \\
        \lstick{$|a_0\rangle$} & \swap{-1} & \qw & \swap{-1} & \qw & \swap{-1} & \qw & \swap{-1} & \qw \rstick{$A\ket{a_0}$}
        \\
    \end{quantikz}
    \caption[Fixed order circuit equivalent to the quantum switch.]{\textbf{Fixed order circuit equivalent to the quantum switch.} In order to implement the same transformation than the quantum switch in a fixed order circuit with unitary inputs $A$ and $B$, at least one extra copy of one of the gates is required. Controlled-SWAP gates are used as a router to ensure the coherent control of the order of the operations.}
    \label{Fig_1_FixedCircuit}
\end{figure}
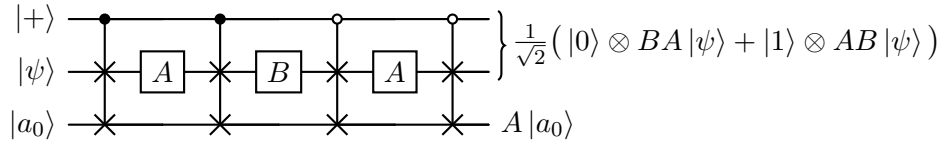

In Ref. \cite{Chiribella_2013} the authors proved that the quantum switch ``cannot be realized by inserting a single use of the input black box in a quantum circuit with fixed causal ordering of the gates''. In other words, if we want to build a quantum circuit implementing this supermap, even in the simplest case of unitary gates, we would need to add at least one additional copy of one of the operations (see Fig. \ref{Fig_1_FixedCircuit}). Furthermore, for general quantum channels it is still unknown whether the quantum switch can be deterministically simulated with a finite number of copies of each channel \cite{Bavaresco_2024}. 
This result strongly suggests that the quantum switch may provide some computational advantages by reducing the number of queries to the quantum operations involved in some particular tasks, and it proves that the quantum switch is an instance of indefinite causal order, what was later confirmed with the definition of causal nonseparability and the study of causal witnesses (cf. Section \ref{sec1_1_sub_CertCausNSep}).

\begin{figure}[t]
    \centering
    \includegraphics[width=0.9\linewidth]{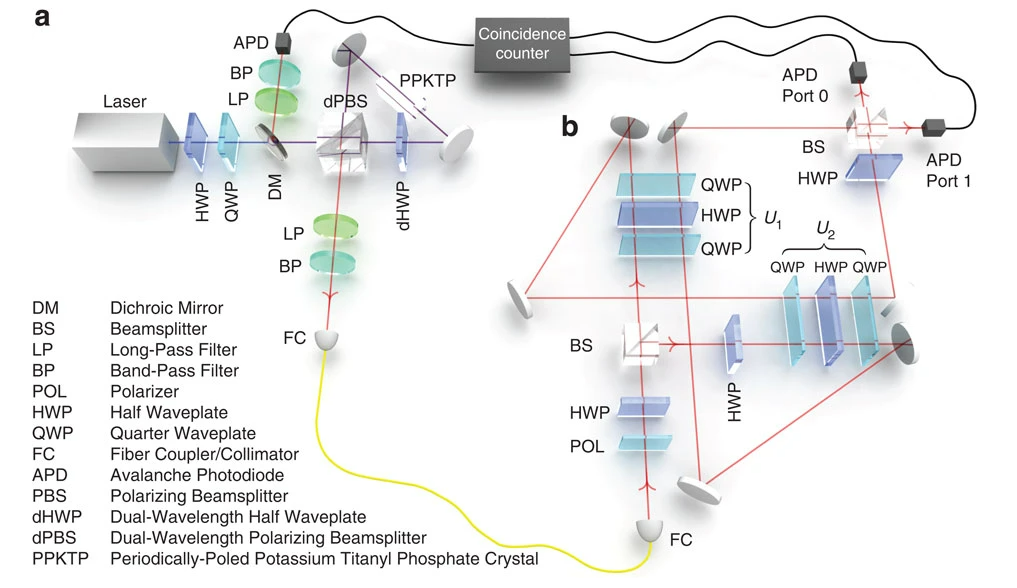}    
    \caption[Photonic quantum switch by Procopio et al. (2015).]{\textbf{Photonic quantum switch by Procopio et al. (2015).} (a) Source of entangled photons. One photon is used as a herald while the other is routed into a quantum switch. (b) Quantum switch. Unitaries $U_1$ and $U_2$, which act on the polarization degree of freedom, are applied on the photon in a superposition of two different orders, coherently controlled by the path degree of freedom. Figure from Ref. \cite{Procopio_2015}.}
    \label{Fig_1_Qswitch_Procopio}
\end{figure}

Despite the impossibility of implementing the quantum switch as a quantum circuit, it is still possible to build it as an interferometric setup. Indeed, as Fig. \ref{Fig_1_Qswitch} suggests, if a given system can be routed through static devices in a superposition of two different paths, then the path followed by the system can act as control and some internal degree of freedom as target. This was the proposal by Procopio et al. in 2015, who reported the first experimental implementation of the quantum switch using a photonic platform \cite{Procopio_2015}. In their experiment, the path of a photon is used as the control system and its polarization as the target. The superposition of paths is achieved using a balanced beam splitter and the operations acting on the polarization degree of freedom are unitary transformations implemented each one as a set of one half-wave and two quarter-wave plates. Then, a second beam splitter allows to measure the control system in the $\sigma_x$ basis (see Fig. \ref{Fig_1_Qswitch_Procopio}). The result of the measurement deterministically discriminates between commuting and anticommuting gates. Following this first experiment, many photonic implementations of the quantum switch have been reported, both for applications and certification of causal nonseparability \cite{Rubino_2017,Goswami_2018,Wei_2019,Goswami_2020,Guo_2020,Rubino_2021,Cao_2022,Rubino_2022,Cao_2023,Liu_2023,Schiansky_2023,Stromberg_2023,Yin_2023,Zhu_2023,An_2024,Antesberger_2024,Li_2024,Tang_2024}. The quantum switch has been also implemented on nuclear spins using nuclear magnetic resonance and coherently controlled interactions \cite{Nie_2022,Xi_2024}. For a thorough review on the experimental implementations of the quantum switch, see Ref. \cite{Rozema_2024}.

A generalization of the quantum switch to a larger number of gates, known as the \textit{$N$-switch}, was introduced by Araújo, Costa and Brukner in 2014 \cite{Araujo_2014}. It consists in the coherent control of the order of $N$ transformations, although it is also possible to use only $p$ of the $N!$ permutations of theses gates \cite{Procopio_2020,Sazim_2021,Wilson_2021}. Following Ref. \cite{Escandon_2023}, we refer to this generalised quantum switch as \textit{$(N,p)$-switch}\footnote{Ref. \cite{Das_2022} introduced a quantum switch of quantum switches, later generalised to higher-order quantum switches, also called \textit{superswitches} \cite{Kechrimparis_2024,Kechrimparis_2025}. These nested quantum switches form a particular class of the \textit{$(N,p)$-switch}.}. The only experimental realization of a quantum switch with more than two gates so far was reported in 2021 by Taddei et al. \cite{Taddei_2021}.

\subsection{Applications}
\label{sec1_1_sub_AppQS}

Since its introduction, the quantum switch has seen a large amount of applications in several areas such as quantum computing, quantum communication, quantum thermodynamics and quantum metrology. Let us revisit some of them in this section.

\paragraph{Quantum computing.}
The first application of the quantum switch outperforming fixed order circuits was proposed by Giulio Chiribella in 2012. He showed that the quantum switch can deterministically discriminates between a pair of commuting and a pair of anticommuting unitaries, while fixed order circuits can only do this probabilistically \cite{Chiribella_2012}. This task was implemented experimentally in Refs. \cite{Procopio_2015,Stromberg_2023} and extended and generalised as a family of promise problems using the $(N,p)$-switch in Refs. \cite{Araujo_2014,Taddei_2021,Renner_2022,Escandon_2023}.

Other application of the quantum switch for quantum computing is the inversion of a unitary evolution. A probabilistic protocol that takes an unknown unitary $U$ and applies $U^\dagger$ on an arbitrary qubit was proposed in Ref. \cite{Trillo_2023} and implemented using the quantum switch in Ref. \cite{Schiansky_2023}. It has been shown that some indefinite causal order processes can achieve greater success probability than any fixed order circuit for unitary inversion and unitary transposition when $k$ copies of the unknown unitary are available, although these processes are different from the quantum switch and there is no clear intuition behind them \cite{Quintino_2019_PRL,Quintino_2019_PRA,Quintino_2022}.

The potential use of indefinite causal order processes for quantum computing has been also studied in connection with nonlocal quantum operations \cite{Ghosal_2023}, Boolean functions \cite{Abbott_2024}, quantum machine learning \cite{Ma_2024}, nonstabilizer operations \cite{Mo_2024} and distributed quantum computing \cite{Liu_2025}.

\paragraph{Quantum communication.}
The first application of the quantum switch for quantum communication was introduced in Ref. \cite{Feix_2015}. There, it was proposed a tripartite communication task in which Charlie must calculate a function of two inputs provided by Alice and Bob. If the communication allowed between the parties is restricted to two bits, then the probability of success using fixed order processes is less than one, while the quantum switch allows the parties to succeed with probability 1. In a subsequent work, the same authors showed a similar task where the quantum switch offers an exponential saving in communication in the asymptotic limit \cite{Guerin_2016}. This advantage was experimentally demonstrated in 2019 in the first implementation of the quantum switch with a high-dimensional target system \cite{Wei_2019}.

The quantum switch admits general quantum channels as input and not only unitary transformations, which are just a particular class of quantum channels. Regarding noisy channels, Ebler, Salek and Chiribella showed in 2018 one of the most striking uses of the quantum switch: the activation of channel capacity. This phenomenon consists in an increase of the amount of information that a quantum channel can transmit. In particular, they showed that if two completely depolarizing channels, which have zero capacity, are applied in a superposition of two different orders, then some amount of information can now be transmitted \cite{Ebler_2018}. The capacity activation mediated by the quantum switch is a phenomenon that appears also in the more general case of $N$ partially depolarizing channels in a superposition of $p\leq N!$ orders \cite{Procopio_2019,Procopio_2020,Procopio_2021,Chiribella_2021_PRL,Wilson_2021,Sazim_2021}. Furthermore, the phenomenon has been studied for general Pauli channels \cite{Delgado_2020,Chiribella_2021_NJP,Das_2022,Kechrimparis_2025}, thermalizing channels \cite{Liu_2022} and amplitude damping channels \cite{Caleffi_2023}. One outstanding result was provided in Ref. \cite{Chiribella_2021_NJP}, where it was shown that two copies of a specific entanglement-breaking channel in a quantum switch exhibit perfect transmission of quantum information if it is heralded by a measurement on the control system.

The first experimental demonstration of the activation of channel capacity was reported in 2020 by two independent groups, both of them implementing two Pauli channels in a photonic quantum switch, although using different degrees of freedom of the photons \cite{Goswami_2020,Guo_2020}. Perfect transmission through noisy channels in a quantum switch was reported in Refs. \cite{Rubino_2021} and the quantum switch of thermal channels was recently implemented in a photonic setup \cite{Tang_2024} and in a nuclear magnetic resonance experiment \cite{Xi_2024}.

\begin{figure}[t]
    \centering
    \begin{tabular}{cc}
        \includegraphics[width=0.5\linewidth]{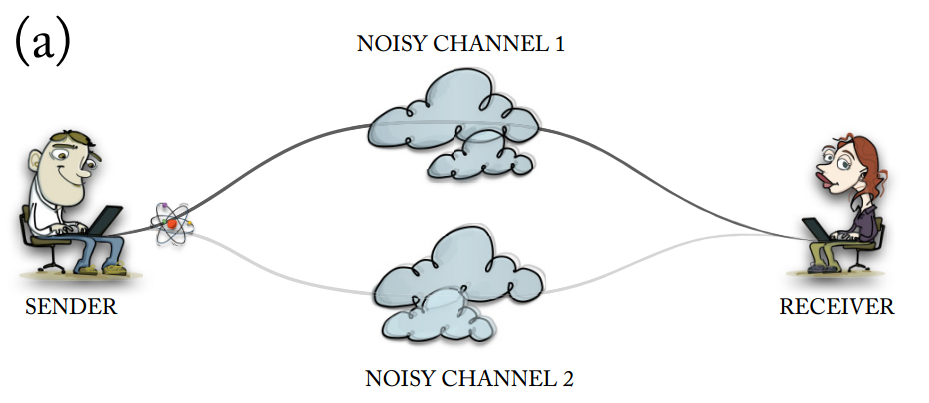}
        &
        \includegraphics[width=0.5\linewidth]{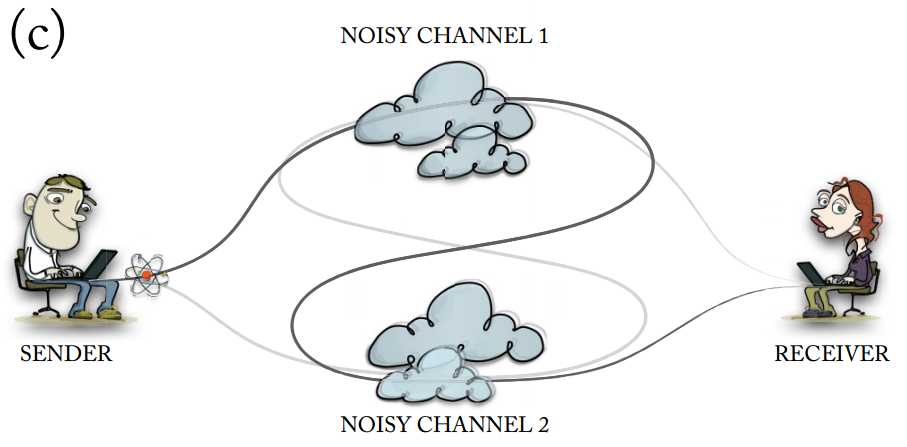}
    \end{tabular}
    
    \caption[Coherent control of operations versus coherent control of orders.]{\textbf{Coherent control of operations versus coherent control of orders.} In the coherent control of operations (left), also called ``superposition of quantum channels'', two quantum channels are applied on a target system, each one in a different arm of an interferometer. Instead, in the quantum switch, or coherent control of order (right), both channels are applied in a sequence whose order depends on the path followed by the target system. Figure from Ref. \cite{Rubino_2021}. The original figure also shows two additional arrays of two quantum channels, not discussed here.}
    \label{Fig_1_SuperpositionOfChannels}
\end{figure}

The capacity activation of noisy channels using the quantum switch is a major result that has been shown to enhance the performance of several communication tasks, such as entanglement distribution \cite{Caleffi_2020,Xu_2023}, teleportation \cite{Mukhopadhyay_2020,Ban_2022,Dey_2025}, superdense coding \cite{Chandra_2022}, quantum random access codes \cite{Mitra_2023}, quantum steering \cite{Mitra_2023}, quantum state discrimination \cite{Kechrimparis_2024} and quantum key distribution \cite{Wu_2025}. Notwithstanding, this effect is not based solely on indefinite causal order. Indeed, Ref. \cite{Chiribella_2019} showed that it is possible to obtain an activation of the channel capacity also in a scenario where two noisy channels are coherently controlled, that is, when each channel is located in a different arm of an interferometer, such that one of them is applied on the target when the control system is in state $\ket{0}$ and the other is applied when the state of the control is $\ket{1}$ (see Fig. \ref{Fig_1_SuperpositionOfChannels}). However, this scenario has subtle differences with the quantum switch. As showed by Abbott et al. in Ref. \cite{Abbott_2020}, additional information regarding the implementation of the quantum channels is required in the coherent control of operations, since the capacity of the effective channel can be different for different implementations of the same noisy channels, while in the quantum switch the effect is independent of their particular implementations. Another difference was highlighted in Ref. \cite{Loizeau_2020}, where it was shown numerically that the coherent control of quantum channels always provide a communication advantage, while the quantum switch can give either an advantage or disadvantage depending on which the input channels are. Interestingly, for some particular cases the quantum switch can achieve perfect transmission, what was proved to be impossible for the coherent control of noisy channels \cite{Chiribella_2021_NJP}. An experimental comparison of these two scenarios was implemented in Ref. \cite{Rubino_2021}. Also, the effect of using different implementations of the same quantum channels with a coherent control was experimentally studied in Ref. \cite{Pang_2023}. It is important to notice that a third scenario was also proposed in Ref. \cite{Guerin_2019}, in which two noisy channels are implemented in a fixed order, but adding controlled unitaries before them, after them and/or in between. Although this third scenario also offers  perfect transmission, as far as we known it has not been further explored. 
All these works are part of an ongoing debate on the role of indefinite causal order in the activation of channel capacity. In some of them it is claimed that the resource that allows capacity activation is the coherence of the control system. 
Indeed, the advantages provided by the quantum switch are affected by decoherence on the control system, as firstly noticed in Ref. \cite{Mitra_2023} and further studied in Ref. \cite{Molitor_2024}. Summarizing, the activation of quantum channels capacity is the result of a combination of three factors: indefinite causal order, coherent control of operations and the particular implementation of the quantum channels.

In addition to this important application of the quantum switch, other communication tasks using the quantum switch have been studied. In particular, entanglement generation with local operations in indefinite order was proposed in Ref. \cite{Koudia_2023}, a protocol for entanglement distillation was devised in Ref. \cite{Zuo_2023} and a key distribution protocol assisted by the quantum switch was studied in Ref. \cite{SpencerWood_2023}. A new cryptographic task called \textit{local-data-hiding} was proposed in Ref. \cite{Naik_2024}, showing that causally nonseparable processes outperform fixed order circuits, although these processes are different from the quantum switch.


\paragraph{Quantum thermodynamics.} 
Another striking application of the quantum switch was suggested by Felce and Vedral in Ref. \cite{Felce_2020}. They considered two copies of a thermalizing channel applied in a superposition of two different orders as in the quantum switch. While the application of these channels in a fixed order always leads to an output thermal state in the same temperature as the thermal baths, the quantum switch allows the target to get a higher or lower temperature than the baths depending on the result of the measurement performed on the control. The authors use this result to build a refrigeration cycle that extracts energy from a cold reservoir while consuming coherence of the control system. This effect was experimentally implemented in 2022 in two independent works, one of them using a photonic setup \cite{Cao_2022} and the other one being the first realization of the quantum switch using nuclear magnetic resonance \cite{Nie_2022}. The effect of decoherence of the control in this task was studied in Ref. \cite{Molitor_2024}.

\paragraph{Quantum metrology.} 
\label{sec1_app_metrology}
Other important avenue for applications of the quantum switch is quantum metrology. One of the first metrological tasks using the quantum switch was proposed in Ref. \cite{Frey_2019}, where the goal was to estimate the strength of a depolarizing channel. To achieve the goal, two copies of the same depolarizing channel are applied on the target system in a superposition of orders. By comparing the quantum Fisher information, it was shown that the strategy using the quantum switch can reach a better precision than strategies applying the operations in a fixed order. A similar idea was also proposed in Ref. \cite{Mukhopadhyay_2018}, where the aim was to measure the temperature of two copies of a thermal bath. These two instances of single parameter estimation using the quantum switch showed that the quantum switch can offer a metrological advantage compared to fixed order strategies.
Further instances of single parameter estimation with enhanced precision via the quantum switch are the estimation of the product of two displacements in the phase space\cite{Zhao_2020,Yin_2023}, estimation of the phase of a noisy unitary gate\cite{Chapeau_Blondeau_2021,Chapeau_Blondeau_2022,Chapeau_Blondeau_2023,An_2024,Yuan_2025}, estimation of the phase of an ideal unitary gate sorrounded by two noisy channels in a superposition of orders \cite{Ban_2023,Ban_2024}, estimation of the rotation angle in an orbital angular momentum state \cite{Li_2024} and estimation of an area on the Bloch sphere \cite{Barnett_2024}. Some of these tasks have been demonstrated in photonic experiments \cite{Yin_2023,An_2024,Li_2024}. A proposal for using the quantum switch for multiparameter estimation, again showing an improvement in the precision of the estimation, was studied in Ref. \cite{Goldberg_2023}.

\paragraph{Others.} 
The list of applications of the quantum switch also includes charging quantum batteries \cite{Zhu_2023}, applying non-gaussian operations \cite{Koudia_2021} and measuring the incompatibility of observables \cite{Gao_2023}. The topic is becoming popular with a larger amount of research being performed in this subject every year.

\section{Process matrices}
\subsection{Concept}
\label{sec1_1_sub_ProcMatr}

\begin{figure}[t]
    \centering
    \includegraphics[width=0.6\linewidth]{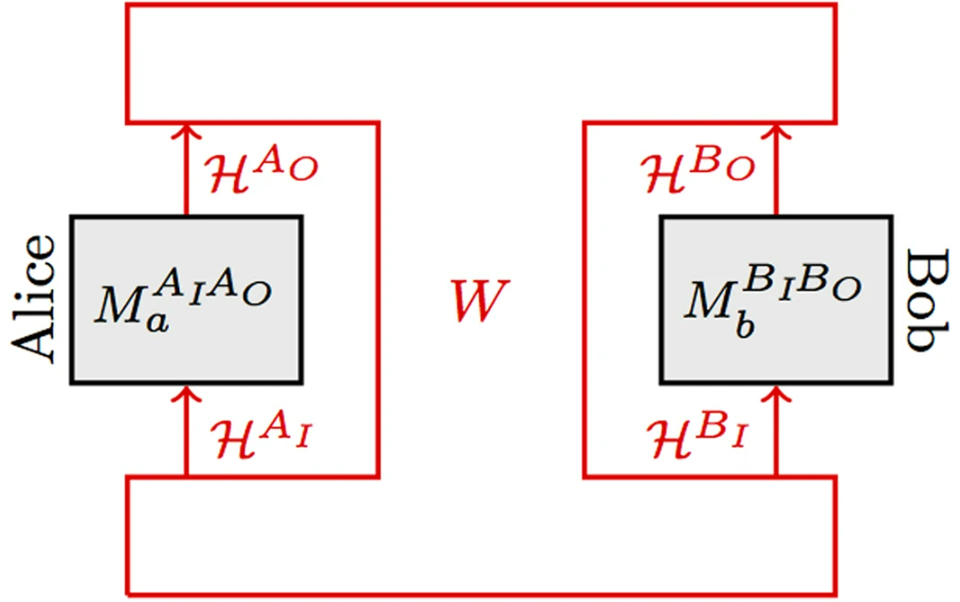}
    \caption[Bipartite process matrix.]{\textbf{Bipartite process matrix.} Alice and Bob perform quantum operations acting on some input systems and obtain outcomes $a$ and $b$. Their labs are connected by an unknown causal structure represented by a process matrix $W$, which may be causally non-separable. The joint probabilities $P(a,b)$ depend on a combination of $W$ and the operations performed by the parties. Figure from Ref. \cite{Branciard_2016_SR}.} 
    \label{Fig_1_ProcessMatrix}
\end{figure}

The quantum switch was introduced following a strictly computational motivation, inspired by the research program of Lucien Hardy. But does quantum mechanics allow other squemes with no definite causal order? Contemporarily to the introduction of the quantum switch, Ref. \cite{Oreshkov_2012} proposed the framework of process matrices, which describes all the correlations that two parties can observe in their laboratories when quantum mechanics is valid within each lab, but without making any assumption about the causal structure out of the labs. Each party, let us call them Alice and Bob, receives a quantum system as input, on which they perform some local quantum operations (see Fig. \ref{Fig_1_ProcessMatrix}). Given a fixed set of operations, the set of joint probabilities that Alice and Bob produce as outcome is called a \textit{process}. Each process is represented by a matrix $W$, which must satisfy some conditions to be a valid process matrix: it must 1) be positive semidefinite in order to ensure that probabilities are non-negative, and 2) satisfy a completeness relation guaranteeing that the sum of all probabilities is $1$. If we also assume that the global causal structure forbids signaling from one of the parties to the other one, then an extra condition must be imposed on $W$, which becomes a channel with memory as defined in Ref. \cite{Chiribella_2009}. A process where Alice cannot signal to Bob, what means that Alice is not in the causal past of Bob, is denoted by $W^{A\nprec B}$. Similarly, $W^{B\nprec A}$ denotes a process where Bob cannot signal to Alice. Using this notation, Ref. \cite{Oreshkov_2012} introduced the notion of \textit{causal non-separability}, formalizing the definition of indefinite causal order. A process $W$ is said to be \textit{causally separable} if it can be written as a convex combination $W = qW^{A\nprec B}+(1-q)W^{B\nprec A}$, where $q\in[0,1]$. Conversely, if $W$ cannot be written in this form, i.e. if it cannot be decomposed as a mixture of definite order processes, it is said to be \textit{causally non-separable}.

In addition to the introduction of the process matrices framework and the definition of causal non-separability, Ref. \cite{Oreshkov_2012} also introduced a particular example of causally non-separable process, later called the ``OCB process''\footnote{``OCB'' refers to the authors of Ref. \cite{Oreshkov_2012}, namely Oreshkov, Costa and Brukner.}. The proof of its causal non-separability consists in the violation of a ``causal inequality'', similar to the violation of a Bell inequality for proving that a given bipartite quantum state is entangled. The proof can be summarized as follows: when Alice and Bob are connected by a causally separable process, their success probability in the ``Guess Your Neighbour Input'' game is always less than $3/4$, while the OCB process allows the parties to exceed this bound; thus, the OCB process must be causally non-separable.

It is important to note that Ref. \cite{Oreshkov_2012} did not offer a way to implement the OCB process, letting open the question about its physicality. Furthermore, it rose the question about which processes in the framework are physically implementable and which are not. This question was addressed in Refs. \cite{Araujo_2017} and \cite{Oreshkov_2019}, which proposed some demarcation criteria for the implementability of general processes. Additionally, Ref. \cite{Purves_2021} showed that the probabilities arising from standard quantum theory do not violate causal inequalities of the type violated by the OCB process, what means that the OCB process cannot be physically implemented according to quantum theory and thus confirms that the process matrices framework goes beyond quantum mechanics. However, the proof in Ref. \cite{Purves_2021} does not consider the possibility of realizing indefinite causal order in time-delocalized subsystems as suggested by Ref. \cite{Oreshkov_2019}; moreover, there exist tripartite processes which admit such a realization and also violate causal inequalities \cite{Wechs_2023}. Thus, the question about the implementability of general processes remains open.

Regarding the quantum switch, it has been shown that it is a causally non-separable process that does not violate any causal inequality \cite{Araujo_2015}. However, and despite its several experimental implementations, there is still an active debate concerning its physicality. Indeed, some authors claim that the reported experiments are just simulations instead of genuine implementations of the quantum switch \cite{Paunkovic_2020,Vilasini_2024}. We are going to postpone this debate for later (cf. Section \ref{sec1_1_sub_GravSwitch}).

In the search for physically implementable processes with indefinite causal order others than the quantum switch, Ref. \cite{Wechs_2021} characterised a broad class of implementable processes called ``Quantum circuits with coherent control''. The quantum switch is an instance of this family, while processes violating causal inequalities, such as the OCB process, cannot be described by it. A novel example of causal non-separable process within this class is a tripartite process that involves dynamical control of the causal order \cite{Wechs_2021}. To the best of our knowledge, this process, known as the ``Grenoble process'', has not been implemented in any experiment so far.

Now let us comment some extensions of the process matrices framework. First of all, the extension to multipartite scenarios was not as straightforward as originally thought by Oreshkov, Costa and Brukner in their seminal work \cite{Oreshkov_2012}. Actually, the multipartite scenario exhibits a richer causal structure than the bipartite case. For example, in the multipartite case the result of an operation performed by one party could influence the order of the operations performed by other two parties \cite{Oreshkov_2016}, a dynamical feature which is absent in the bipartite case. Also, an $N$-partite causally non-separable process is not necessarily genuinely $N$-partite non-causal, since the parties could eventually be grouped in smaller sets with a definite causal order between the sets \cite{Abbott_2017}. Another difference between the bipartite and multipartite cases relates to the restriction to classical parties. While all the correlations between two parties that only perform classical operations can be explained by causally separable processes \cite{Oreshkov_2012}, this is not true in the multipartite case \cite{Baumeler_2014,Baumeler_2016}. For a discussion on the definition of multipartite causal nonseparability, see Ref. \cite{Wechs_2019}.

The process matrices formalism and its multipartite extension was introduced originally for finite dimensional systems. Its continuous-variable (CV) version, i.e. with parties performing operations on CV quantum systems, was studied in Ref. \cite{Giacomini_2016}, but only for the bipartite case. That work showed that the quantum switch is a causal non-separable process also in the CV regime. However, to our knowledge, no certification strategy has been proposed for this scenario. The extension of the process matrices framework to the CV multipartite case, with its expected richer structure, is lacking as well.

Finally, we can mention a few further developments on the process matrices framework. For instance, Ref. \cite{Guerin_2019_NJP} studied the possibility of composing process matrices, Ref. \cite{Lewandowska_2023} addressed the discrimination of process matrices and Ref. \cite{Antesberger_2024} proposed and implemented a scheme for process matrix tomography. The possibility of process matrices evolving in time was studied in Ref. \cite{Selby_2024}. Lastly, Refs. \cite{Guerin_2018,Wechs_2024} studied the so called ``causal reference frames'', i.e., how a process is observed from the perspective of each party involved in the process.

\subsection{Certification of causal non-separability}
\label{sec1_1_sub_CertCausNSep}

How can we certify that a given quantum process or experiment is an instance of indefinite causal order? In this subsection we are going to take a glance at the different strategies that answer this question, which mimic the methods for entanglement certification.

We mentioned above that some causally non-separable processes violate some causal inequalities. Causal inequalities allow for a device-independent certification of indefinite causal order, since they, as well as Bell inequalities, rely solely on the outcome probabilities of the operations performed by each party involved in the process. In the case of entanglement, tight Bell inequalities are the facets of the ``Local Polytope'', wich is the set of probabilities compatible with local hidden variable models \cite{Guhne_2009}. In analogy to this description, Ref. \cite{Branciard_2016_NJP} characterized the ``Causal Polytope'' for bipartite processes, that is the set of correlations compatible with definite causal order. The facets of the causal polytope define causal inequalities which can be violated for some causally non-separable processes. The causal polytope for the tripartite scenario was characterised in Ref. \cite{Abbott_2016}. In entanglement certification, the extent to which a Bell inequality can be violated has been shown to have an upper limit given by the Tsirelson bound \cite{Cirelson_1980}. Similarly, a Tsirelson-like bound for the violation of causal inequalities was shown in Refs. \cite{Brukner_2015,Liu_2024}. We have also mentioned above that there exist causally non-separable processes, like the quantum switch, that do not violate causal inequalities. Notwithstanding, by adding an extra party it is possible to build some inequalities that can be violated by the quantum switch. For instance, Ref. \cite{vanderLugt_2023} proposed an inequality for a quantum switch whose control is entangled with a qubit sent to a space-like separated party who performs measurements on it. This proposal shows that device-independent certification of indefinite causal order is possible. 

\begin{figure}[t]
    \centering
    \includegraphics[width=0.8\linewidth]{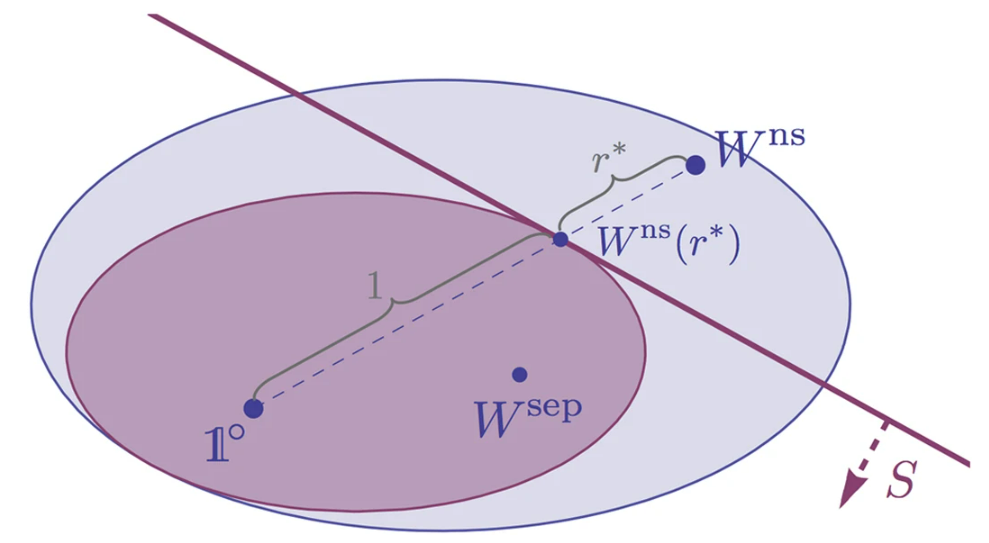}
    \caption[Causal Witness.]{\textbf{Causal Witness.} The inner ellipse represents the set of causally separable processes and the bigger ellipse represents the set of all process matrices. For every causally non-separable process $W^{ns}$, it can be defined a hyperplane $S$ (a causal witness) that separtates $W^{ns}$ from the set of causally separable processes. Figure from Ref. \cite{Branciard_2016_SR}.} 
    \label{Fig_1_CausalWitness}
\end{figure}

One of the most important methods to certify entanglement is the measurement of entanglement witnessess \cite{Guhne_2009}. For different quantum states, different witnesses must be build, and their implementation requires prior knowledge about the particular state and the functioning of the devices used by the parties; hence, it is a device-dependent certification strategy. Again, it is possible to mimic this method and build ``causal witnesses'' for certifiyng causal non-separability \cite{Araujo_2015,Branciard_2016_SR}. Indeed, given that the set of causally separable processes is a convex set, it is possible to define an hyperplane separating this set from the specific causally nonseparable process to be certified (see Fig. \ref{Fig_1_CausalWitness}). The main advantage of this approach is that a causal witness can be translated into a number of operational settings to be implemented by the parties. In particular, a causal witness for the quantum switch was derived using semidefinite programming in Ref. \cite{Araujo_2015} and Refs. \cite{Rubino_2017} and \cite{Goswami_2018} independently demonstrated that the quantum switch is an instance of indefinite causal order by measuring a causal witness in photonic experiments.

An intermediate approach, known as semi-device-independent certification of causal non-separability, was formulated in Ref. \cite{Bavaresco_2019}. There, one of the parties is trusted, meaning that their operations are fully characterised, while the other one is untrusted and only their outcome probabilities are known. Correlations certified this way are stronger than those certified only by device-dependent strategies. The causal non-separability of the quantum switch can be certified using this method \cite{Bavaresco_2019} and has been demonstrated experimentally \cite{Cao_2023}. An alternative semi-device-independent method was proposed in Ref. \cite{Dourdent_2022}, where all the parties remain uncharachterised, but they are given trusted quantum inputs instead of classical ones. By adding new uncharacterised parties, although connected in a particular network, this semi-device-independent with quantum inputs strategy becomes a network-device-independent certification of causal non-separability. \cite{Dourdent_2024}.

Lastly, a possibilistic proof of causal non-separability of the quantum switch has been recently introduced \cite{vanderLugt_2024} in analogy to the possibilistic proof of nonlocality by Peres and Mermin \cite{Peres_1990,Mermin_1990}.

Summarizing, different tools used for proving nonlocality of quantum states, both probabilistic and possibilistic, have been translated for certifying causal non-separability of process matrices. This spectrum of strategies also shows that quantum processes can be sorted in a hierarchy with different causal non-separability strengths, although, to our knowledge, no measure of causal non-separability has been proposed within this framework.

\section{Superposition of causal structures}
\label{sec1_1_sub_GravSwitch}

The purpose of the Hardy's program is to build a theory of quantum gravity, but neither the quantum switch nor the process matrices framework deal with gravity explicitly. Alternatively, Ref. \cite{Zych_2019} put features from both gravity and quantum theory together, by assuming the possibility of a superposition of causal structures. The idea is the following: let us suppose that each metric defining a manifold $\mathcal{M}$ can be assigned a quantum state $\ket{\mathcal{M}}$, with orthogonal states representing macroscopically distinguishable spacetimes; then, we assume that a superposition of different manifolds, such as $(\ket{\mathcal{M}_1}+\ket{\mathcal{M}_2})/\sqrt{2}$, is a valid state for the geometry of spacetime. We could guess that a theory of quantum gravity would eventually admit this kind of composition. If so, what kind of phenomena should we expect from such a superposition of causal structures?

\begin{figure}[t]
    \centering
    \includegraphics[width=0.8\linewidth]{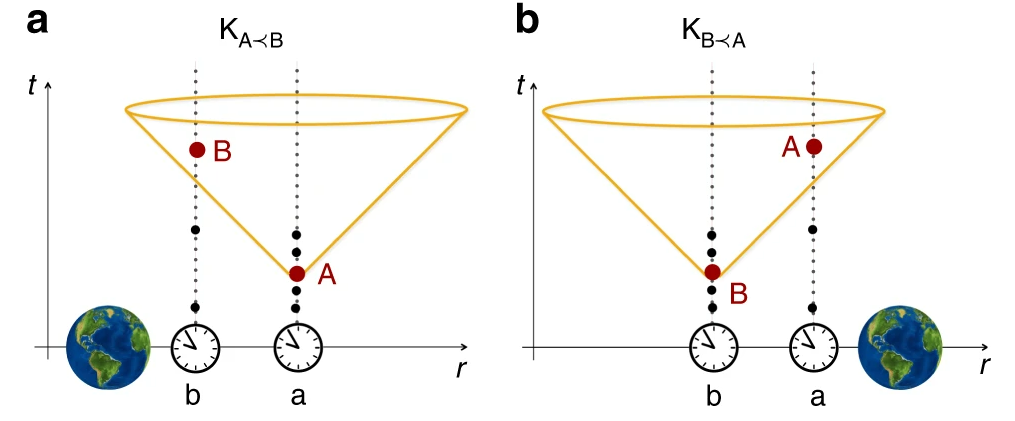}
    \caption[Gravitational quantum switch.]{\textbf{Gravitational quantum switch.} A massive body is in a superposition of two different locations, while two clocks are in fixed locations. Due to gravitational time dilation, an event defined in one clock can be in the causal past of an event defined in the other clock. The order of the events depends on the branch of the superposition; an indefinite location of the massive body generates an indefinite causal structure realising a gravitational version of the quantum switch. Figure from Ref. \cite{Zych_2019}.}
    \label{Fig_1_GravSwitch}
\end{figure}

Ref. \cite{Zych_2019} studied a particular scenario where a massive body is in a superposition of two different locations. There, the gravitational field is naturally in a superposition of two different geometries. As seen in Fig. \ref{Fig_1_GravSwitch}, we could locate two clocks in fixed positions respect to a far-away observer, such that in one configuration one of the clocks will be closer to the massive body, while in the other branch of the superposition the other clock will be the one closer to the source of the gravitational field. For a given proper time $\tau$ the clocks define corresponding events $A$ and $B$. However, for each configuration, if the clocks start synchronised as seen from the far-away observer, after a while their proper times will be different due to gravitational time dilation. Hence, one of the events will be in the causal past of the other. For the superposition of the two geometries, the result will be a superposition of order of the events, as in the quantum switch. Moreover, the quantum switch can be implemented in this scenario by locating a target system together with one of the clocks and applying on it local quantum operations triggered by a signal emmited from each clock at its proper time $\tau$. This version of the quantum switch, where the control system is the gravitational field prepared in a superposition of classical manifolds, is usually called a \textit{gravitational quantum switch}. An alternative realization of the gravitational quantum switch, where the source of the gravitational field is a superposition of spherical mass shells, was proposed in Ref. \cite{Moller_2024}.

Notice that in the scenario described in the previous paragraph there are two parties or labs with their own clocks either applying or triggering quantum operations on a target system. According to the reference frame of each lab there is one quantum operation applied in a definite time while the other operation appears to be time-delocalised. This observation led to the concept of ``time reference frames'', i.e., quantum reference frames associated to gravitating clocks, studied in Ref. \cite{CastroRuiz_2020}. Other realizations of the gravitational quantum switch consider agents in a superposition of paths \cite{Moller_2021} or with entangled acceleration \cite{Dimic_2020} in a fixed spacetime, although these scenarios are different to the superposition of causal structures that we are reviewing here.

\begin{figure}[t]
    \centering
    \includegraphics[width=0.8\linewidth]{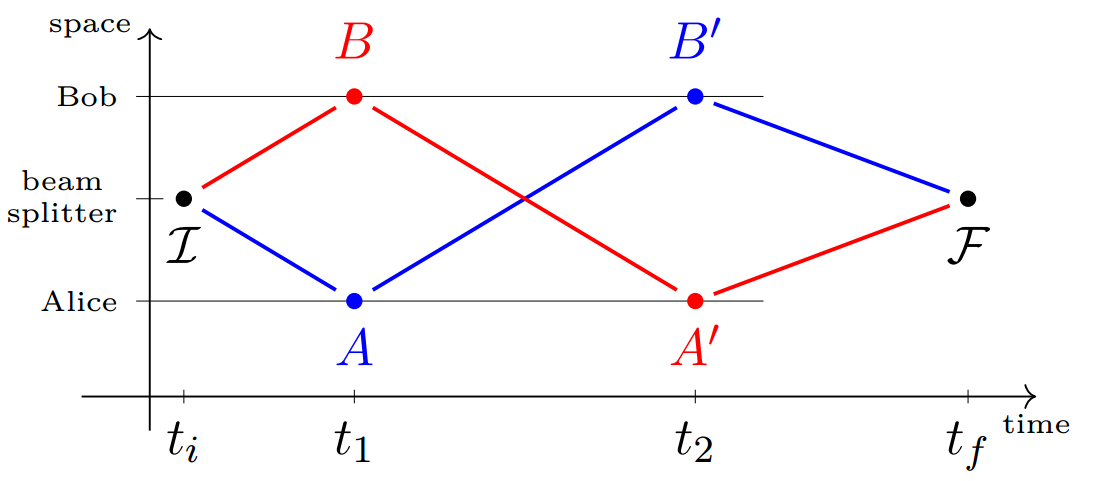}
    \caption[Photonic quantum switch embedded in spacetime.]{\textbf{Photonic quantum switch embedded in spacetime.} The implementation of the photonic quantum switch comprises four points in spacetime. A genuine quantum switch should comprise only two events. Figure from Ref. \cite{Paunkovic_2020}.}
    \label{Fig_1_Switch_spacetime}
\end{figure}

In Section \ref{sec1_1_sub_ProcMatr} we mentioned that some authors claim that the photonic experiments implementing the quantum switch are just simulations of this process \cite{Paunkovic_2020}. According to them, only the gravitational quantum switch can be a genuine implementation of the quantum switch. The reason for that comes from the fact that all the photonic experiments implementing the quantum switch are performed within a laboratory immersed in a fixed Minkowski spacetime. As shown in Fig. \ref{Fig_1_Switch_spacetime}, in a photonic experiment the target system is split in a superposition of paths, with one of them reaching Alice's lab at time $t_1$ and then going to Bob's lab at time $t_2$, while in the other arm of this array the photon goes first to Bob's lab at time $t_1$ and then to Alice's in time $t_2$. Strictly speaking, there are two controlled operations ($A$ and $A'$ in Fig. \ref{Fig_1_Switch_spacetime}), which happen to be equal, performed in Alice's side at different times. The same occurs at Bob's. Thus, the photonic implentation of the quantum switch is just a simulation, since it considers four events in spacetime instead of two as a genuine quantum switch, like the gravitational one, would do. Four controlled operations are explicitly applied also in nuclear magnetic resonance implementations of the quantum switch. 

A different analysis leading to a similar conclusion was carried out in Refs. \cite{Vilasini_2024,Vilasini_2024_PRA}, where it was discussed that there are two different notions of causality involved in the study of the quantum switch. One of them is an information-theoretic concept of causality and the other one is spacetime causality. The latter demands an acyclic spacetime structure in order to satisfy relativistic causality principles such as the impossibility of superluminal signaling. By introducing the concept of ``fine-graining'', Refs. \cite{Vilasini_2024,Vilasini_2024_PRA} showed that cyclic information-theoretic structures are embedded in acyclic spacetime structures, reconciling both notions. Furthermore, they showed that causally nonseparable processes are compatible with cyclic causal structures, therefore they admit a realization in an acyclic spacetime at a fine-grained level as fixed order processes. In particular, since the photonic experiments of the quantum switch are implemented in a definite Minkowski spacetime (acyclic structure satisfying relativistic causality), they must be instances of a fixed order process. The question about how to make general processes compatible with spacetime in the fine-grained level was studied in Ref. \cite{Salzger_2024}.

\begin{figure}[t]
    \centering
    \includegraphics[width=0.95\linewidth]{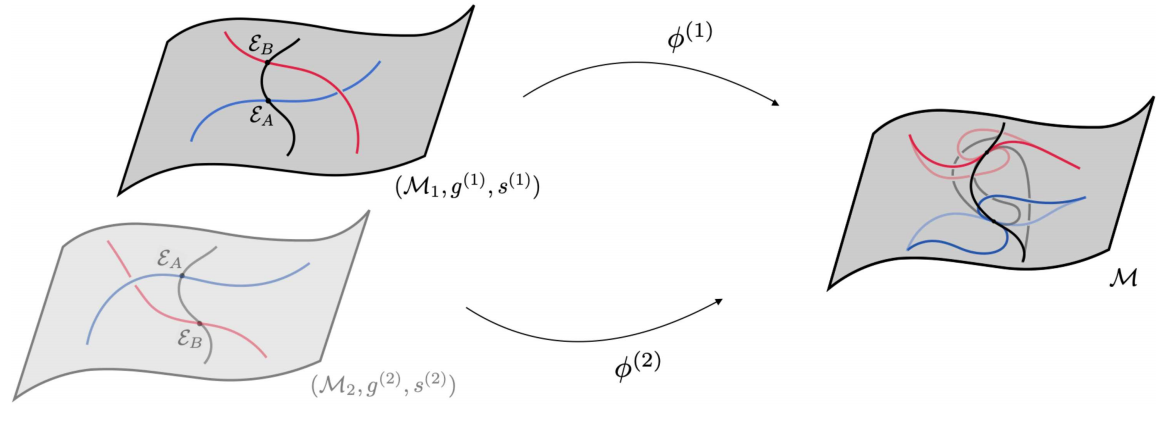}
    \caption[Superposition of causal structures.]{\textbf{Superposition of causal structures.} In the superposition of classical manifolds, an event is a crossing of worldlines. For each branch, the corresponding events are identified via a quantum controlled diffeomorphism. Figure from Ref. \cite{delaHamette_2022}.} 
    \label{Fig_1_Superposition_manifolds}
\end{figure}

Nevertheless, the criticism against a faithful realisation of the quantum switch in photonic experiments strongly relies on the notion of \textit{event} understood as a \textit{point in spacetime}. Although the concept is uncontroversial when the spacetime is definite, it must be revisited when a superposition of causal structures is considered. Indeed, Ref. \cite{delaHamette_2022} advocates for a rather operational definition of an event as a \textit{coincidence of worldlines}, notion further discussed in Ref. \cite{delaHamette_2024}. For each branch of the superposition of manifolds (see Fig. \ref{Fig_1_Superposition_manifolds}), it is possible to draw the worldlines of the target system and the agents Alice and Bob. The application of a quantum operation on the target occurs when the target's worldline meets the worldine of one of the agents. However, since there is a superposition of causal structures, the events of each branch must be identified. This identification can be achieved via a quantum controlled diffeomorphism \cite{delaHamette_2022}. According to this operational view, the photonic implementation of the quantum switch is as genuine as its gravitational version, provided a superposition of causal structures is involved.

The superposition of causal structures is still a new approach with room for further developments. We can mention a few works on this line. For example, some quantifiers for indefinite causal order have been defined in Refs. \cite{delaHamette_2022,Fedida_2024}. Besides, Ref. \cite{Foo_2023_PRD} studied the operational effects of a superposition of Minkowski spacetimes with periodic boundary conditions while Ref. \cite{Liu_2024_Diamond} analyses the influence of a superposition of diamond spacetimes on quantum entanglement. Also, it has been shown that superposition of spacetimes related by coordinates transformations can be expressed as a single definite spacetime with matter in a superposition of configurations \cite{Foo_2023_arxiv}. Actually, this is the case of the scenario considered in Ref. \cite{Zych_2019} with a massive body in a superposition of two locations and two clocks in fixed positions, which is equivalent to a massive body in a fixed position and two clocks with entangled positions. Anyway, there are much more questions to solve in this approach. For instance, if quantum states are assigned to classical manifolds, in which Hilbert space do those states live? How do they evolve? How can the state of a classical manifold be projected onto a superposition of geometries or viceversa? How to identify events when two worldlines meet in one of the branches but not in the other one? As we mentioned earlier, while the superposition of causal structures is a rather speculative approach, it is an interesting avenue to explore new phenomena that could eventually appear in a theory of quantum gravity and therefore in regimes where both quantum and gravitational effects are relevant.

\section{Ongoing debates}
\label{sec1_1_sub_Debates}

We have seen that indefinite causal order is a research topic that embraces from foundational research to applications, linking quantum gravity with quantum information communities. Different approaches have been explored in the last decade in order to formalize the concept and a number of experiments have been already implemented. However, there are still major ongoing debates, which we summarize here.

\begin{enumerate}
\item \textbf{On the advantages of the quantum switch:} The quantum switch offers interesting advantages when compared to fixed order circuits. However, it is not completely clear which is the resource behind those advantages. Is it indefinite causal order or the coherent control of operations? Or maybe is it the non-commutativity of operators? Or a mixture of all of them? Besides, some of the advantages have been shown either in \textit{ad hoc} tasks for the quantum switch or in impractical scenarios. Can they be transferred to real-world problems in order to provide practical solutions?
\item \textbf{On the physicality of process matrices:} The process matrices framework includes processes that violate causal inequalities. However, they seem to be non-physical processes. Which is the boundary between physically implementable processes and non-physical ones? Can those processes forbidden by standard quantum mechanics be relevant processes in a quantum gravity scenario? In other words, can some of them be still allowed by Nature in some extreme conditions?
\item \textbf{On the faithfulness of ICO realisations:} The notion of event is still controversial and it is the key when interpreting current experiments on indefinite causal order. Are the experiments realising the quantum switch genuine instances of indefinite causal order or are they just simulations instead?
\item \textbf{On the definition of quasi-classical manifolds:} The superposition of causal structures is an approach that needs to be further developed. Is the Hilbert space of the states assigned to classical manifolds well defined? How much more structure from quantum theory can be added to this approach?
\end{enumerate}

All these questions are quite interesting, and there are more open questions on each of the specific topics reviewed above. The road is open and new unknown lands are waiting to be explored. Maybe this map can help some enthusiast explorer to find their way, or become an invitation for the curious newcomer. 


\section*{Acknowledgements}

This work was supported by ANID - Subdirecci\'on de Capital Humano/Doctorado Nacional/2021-21211347 and ANID - Millennium Science Initiative Program - ICN17\_012.

\renewcommand\refname{References}          
{\setstretch{1.0}                           
\printbibliography[heading=bibintoc, title=References]
}

\end{document}